\documentclass[aps,twocolumn,floatfix,superscriptaddress,preprintnumbers]{revtex4}
\usepackage{amsfonts}
\usepackage{amssymb}
\usepackage{amsmath,mathtools}

\RequirePackage{ifpdf}
\ifpdf
  \usepackage[pdftex]{graphicx}
\else
  \usepackage[dvipdfmx]{graphicx}
\fi

\renewcommand{\vec}[1]{\mathbf{#1}}

\begin{document}
\preprint{OU-HET-1003}

\title{AdS/CFT as a deep Boltzmann machine}
\author{Koji Hashimoto}
\affiliation{Department of Physics, Osaka University, Toyonaka, Osaka 560-0043, Japan}
%%%%%%%%%
\date{\today}

\begin{abstract}
We provide a deep Boltzmann machine (DBM) for the AdS/CFT correspondence. 
Under the philosophy that the bulk spacetime is a neural network, we give a dictionary between those,
and obtain a restricted DBM as a discretized bulk scalar field theory in curved geometries. 
The probability distribution as training data is the generating functional of the boundary quantum field theory,
and it trains neural network weights which are the metric of the bulk geometry.
The deepest layer implements black hole horizons, 
and an employed regularization for the weights is an Einstein action.
A large $N_c$ limit in holography reduces the DBM to a folded feed-forward architecture. 
We also neurally implement holographic renormalization into an autoencoder.
The DBM for the AdS/CFT may 
serve as a platform for studying mechanisms of spacetime emergence in holography.

%Here, abstract.
\end{abstract}


\maketitle

\setcounter{footnote}{0}

\noindent
%%%%%%%%%%%%%%%%%%%%%%%%%%%%%%
\section{Introduction}

Deep Boltzmann machines \cite{dbm}
are a particular type of neural networks in deep learning \cite{Hinton,Bengio,LeCun} 
for modeling probabilistic distribution
of data sets. They are equipped with deep layers of units in their neural network architecture,
and are a generalization of Boltzmann machines \cite{BMref} which 
are one of the fundamental models of neural networks.
Deepening the architecture enlarges the
representation power of the models, and 
recent advances in training deep models in machine learning 
were initiated by analogues of the deep Boltzmann machines.

The neural network of a deep Boltzmann machine consists of visible units and hidden units. On those
units binary variables live, and they interact with each other under a Hamiltonian called an energy function.
Thus basically the deep Boltzmann machine is an Ising model in which spins only at a boundary layer are
visible (observable), and the Hamiltonian allows inhomogeneity and nonlocality.
For a given probability distribution of the observed spin configurations at the boundary layer,
Ising bond strengths (called ``weights") 
in the model Hamiltonian are trained to approximate the given distribution; that is
the deep learning of the deep Boltzmann machine. 
The training determines the weights automatically, and a structure of the Hamiltonian emerges.
Efficient algorithms  for the training \cite{EfficientDBM} accelerated the progress in deep learning.

\begin{figure}[t]
\begin{center}
\includegraphics[width=6.5cm]{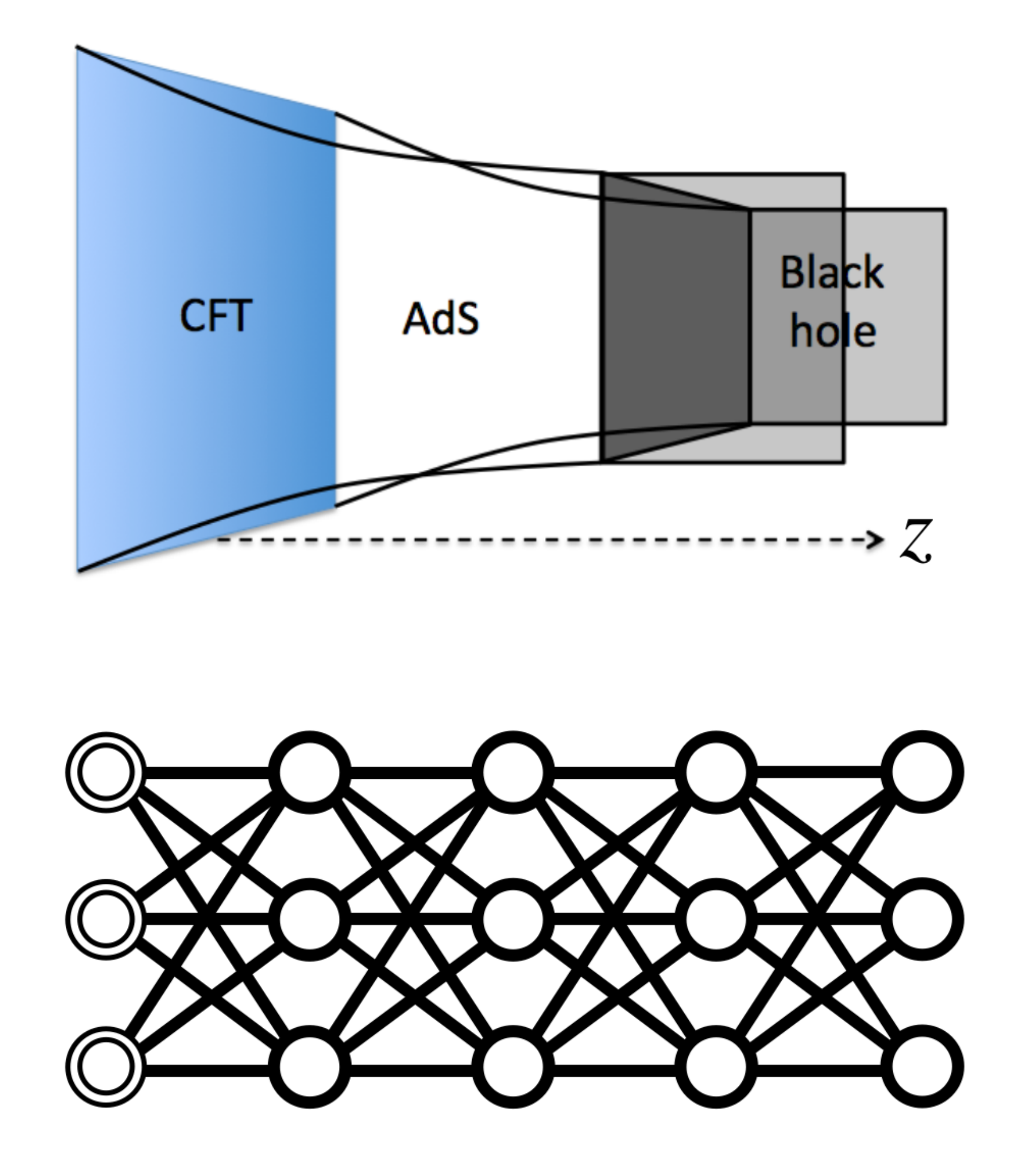}
\caption{
\label{fig:res}
Top: the AdS/CFT correspondence. The horizontal coordinate $z$ is the emergent spatial direction.
Bottom: a deep Boltzmann machine. Circles are units and lines are weights. Double circles are
visible units. It has a layered structure, and ``deep" means that they have many layers toword the
right direction in the figure.
}
\end{center}
\end{figure}

In this paper we study a relation between the deep Boltzmann machines and the AdS/CFT correspondence
\cite{Maldacena:1997re,Gubser:1998bc,Witten:1998qj} 
in quantum gravity. The AdS/CFT correspondence is a holographic duality between 
a $(d+1)$-dimensional quantum gravity and a $d$-dimensional quantum field theory (QFT) without gravity.
The latter lives at the boundary of the gravitational spacetime of the former. From the viewpoint of the QFT,
the direction perpendicular to the boundary surface is an ``emergent" space direction.
Therefore, the aforementioned structure of the deep Boltzmann machines suits the scheme of the AdS/CFT,
once we identify their visible layers with the QFT, and the hidden layers as the bulk spacetime.
See Fig.~\ref{fig:res}.
The trained weights are interpreted as the metric function of the bulk geometry.
We detail the relation between the two schemes both of which are renowned independently 
in different sciences.

%%%%%%%%%%%%%%%%%%%%%%%
%%%%%%%%%%%%%%%%%%%%%%%

A motivation to bring them together also comes from recent progress in discretization of the AdS/CFT.
Popular toy models of the AdS/CFT {\it a la} quantum information use MERA \cite{Swingle:2009bg} 
and other tensor networks \cite{Pastawski:2015qua}. 
In the first place, quantum gravity has a long history of Regge calculus \cite{Regge:1961px}
and dynamical triangulation \cite{Ambjorn:1998xu} where spacetimes are approximated by networks.
For formulating quantum gravity, we need dynamical network whose structure is determined
in a self-organized manner. In that view, neural network architecture may provide a novel platform for
quantum gravity and the emergent spacetime.

We show that the AdS/CFT correspondence naturally fits the scheme of the deep Boltzmann machines, 
where the bulk spacetime geometry is reinterpreted as a sparse neural network.
We construct explicitly a deep Boltzmann machine architecture which represents an example of
the AdS/CFT correspondence.
Previously,  in \cite{Gan:2017nyt, Howard} a possibility of relating hidden variables of Boltzmann machines
to bulk fields was mentioned \footnote{See also an essay \cite{Lee:2017skk}.}.
In \cite{You:2017guh}, entanglement feature of  a free fermion chain was trained at
a random tensor network as a deep Boltzmann machine.
The holographic interpretation for feed-foward deep neural networks 
was proposed and studied in \cite{Hashimoto:2018ftp,Hashimoto:2018bnb} 
for training QFT and QCD linear response functions, and 
the obtained emergent bulk spacetime for large $N_c$  QCD exhibits interesting physical properties
and computes other observables as predictions. While our work here naturally relates to these work,
in this paper we concentrate on deep Boltzmann machines as an AdS/CFT correspondence.

Although at first sight the two schemes look similar, in details they possess different characteristics.
For example, 
since deep Boltzmann machines have constraints for their architecture and 
trainability, we need a careful discretization of bulk field theories.
In addition, the AdS/CFT correspondence is well-understood at the large $N_c$ limit, while
that limit has not been studied in the Boltzmann machines. 
Furthermore, generalization in deep learning owes to degenerate sets of trained weights,
while in the AdS/CFT that degeneration is not expected.
In this paper we address these basic questions raised in relating the two schemes. We provide 
a concrete expression of a deep Boltzmann machine which satisfies the standard constraints,
and find that the large $N_c$ limit brings the Boltzmann machine to a folded feed-forward architecture.
We propose an Einstein action as a regularization of training to distinguish sets of weights to be 
interpreted as a smooth spacetime.

%%%%%%%%%%%%%%%%%%%%%%%
\begin{figure*}
\includegraphics[width=15cm]{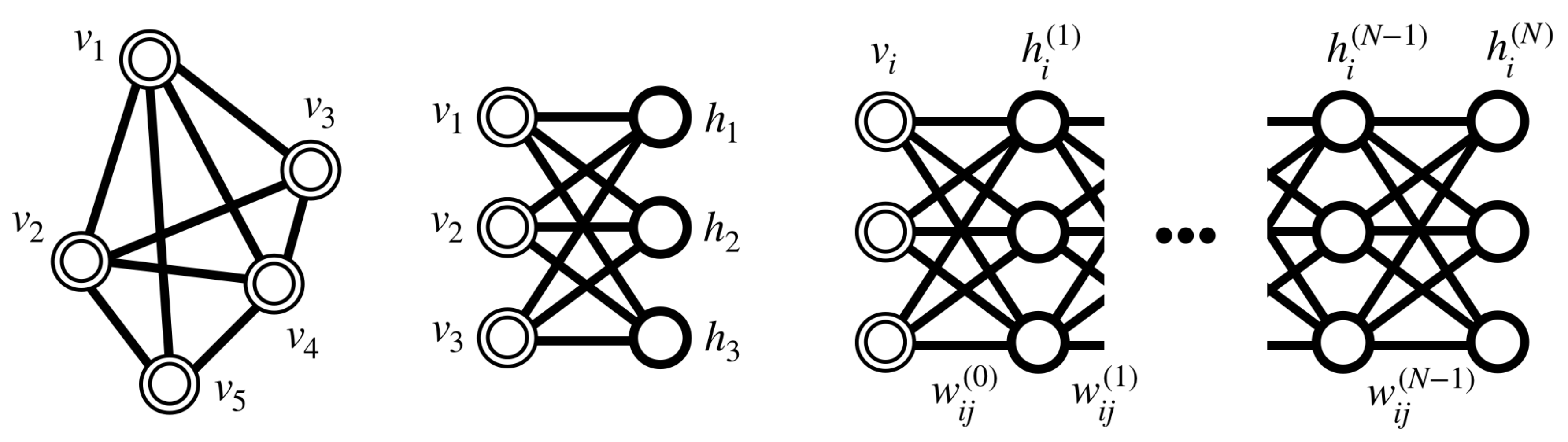}
\caption{
\label{fig:bm}
A schematic picture of Boltzmann machines. Double circles are visible units, and single circles
are hidden units.
 Left: Boltzmann machine. Center: Restricted Boltzmann machine. Right: Deep Boltzmann machine.
}
\end{figure*}
%%%%%%%%%%%%%%%%%%%%%%%

The organization of this paper is as follows. In Sec.~\ref{sec:review}, we briefly review 
deep Boltzmann machines. In Sec.~\ref{sec:holo}, we provide a dictionary of the 
deep Boltzmann machines and the AdS/CFT, and construct a deep Boltzmann machine
for a bulk scalar field theory in generic curved geometry. Discretization of the fields and the
spacetime, and also the properties at the deepest layer are studied. 
In Sec.~\ref{sec:saddle},
we apply the standard large $N_c$ limit (saddle point approximation) to the deep Boltzmann machine
and see the consistency with the holographic linear response.
In Sec.~\ref{sec:Ein}, we propose how we identify weights interpreted as a spacetime,
through a regularization using Einstein action.
Sec.~\ref{sec:Sum} is devoted to our summary and discussions. In Appendix \ref{sec:Auto},
we provide an autoencoder-like neural network architecture for holographic renormalization.

%%%%%%%%%%%%%%%%%%%%%%%%%%%%%%

\section{Brief review of Boltzmann machine}
\label{sec:review}

Boltzmann machines in machine learning are a network model for giving
a probabilistic distribution $P(v_i)$ of the input variables $v_i\in \{0,1\}$ for
$i=1,\cdots,n$. The probability $P(v_i)$ of the Boltzmann machine is defined by
\begin{align}
P(v_i) = \exp\left[- {\cal E}(v_i) \right]
\end{align}
with an energy function ${\cal E}(v_i)$ given by
\begin{align}
{\cal E}(v_i)\equiv \sum_{i} a_i v_i + \sum_{i\neq j} w_{ij} v_i v_j. 
\label{BME}
\end{align}
Here $a_i$ and $w_{ij}$ are real parameters, and are called bias and weight, respectively.
The structure of the Boltzmann machine is specified by a network graph (see Fig.~\ref{fig:bm} Left)
in which $n$ units each of which takes the value $v_i$ 
are denoted by circles while the weights $w$ are
by lines connecting the circles. Obviously, in physics $P$ is a Boltzmann distribution 
of a canonical ensemble of a classical Ising model, after which the name Boltzmann machine 
was named.

As \eqref{BME} is quadratic in $v_i$, even if one varies the biases and the weights,
the class of the probability distributions obtained by \eqref{BME} is quite 
limited. Making the network deep
enlarges the representation power of the architecture. Adding hidden variables $h_i$, we have 
\begin{align}
P(v_i) =\sum_{h_i\in\{0,1\}}
\exp \left[ -{\cal E}(v_i,h_i)\right]
\label{sumh}
\end{align}
with the energy function
\begin{align}
{\cal E}(v_i,h_i)\equiv \sum_{i} (a_i v_i + b_i h_i) + \sum_{ij} w_{ij} v_i h_j \, .
\label{rbm}
\end{align}
The layer consisting of $v_i$ ($h_i$) is called visible (hidden) layer, and 
here, weights connecting units in the same layer are set to zero, so the network graph is restricted to have only
limited connection lines (see Fig.~\ref{fig:bm} Center). This \eqref{rbm} is called
a restricted Boltzmann machine. 

%This restriction is important for training the machine. 
Suppose one has measured events and obtained 
many sets of $\{v_i\}$. From that one can calculate a statistical probability distribution 
$P_{\rm ev}(v_i)$. 
In machine learning, one trains the machine to mimic $P_{\rm ev}(v_i)$. 
In the function $P(v_i)$ of the Boltzmann machine, the weights and biases are trained parameters.
The difference measure between the two probability distributions, called error function,  is the relative entropy
(alternatively called Kullback-Leibler (KL) divergence in machine learning)
\begin{align}
D_{\rm KL}\left(P_{\rm ev}(v_i) \big|\big| P(v_i)\right) 
\equiv \sum_{\{ v_i \}}P_{\rm ev}(v_i) \log \frac{P_{\rm ev}(v_i)}{P(v_i)}
\label{KLPP}
\end{align}
and one tries to minimize it by changing the parameters. When this divergence is minimized,
the machine is well-trained. 

The reason why the restriction in the graph of the Boltzmann machines is important is
the conditional independence. In \eqref{rbm}, when $\{ h_i\}$ is given, 
${\cal E}$ is linear in $v_i$, so the probability $P$ factorizes to a product of each unit $v_i$,
then the training requires a lot less computational resource.
Note that without losing this conditional independence we can add a term 
$\sum_{i,j} w_i \delta_i^j h_i h_j$;
The Kronecker delta $\delta_i^j$ means that this is a self-interaction within the same unit.
What is not allowed for the conditional independence is the term like $h_i h_j$ or $v_i v_j$
with $i\neq j$
in the same layer.

Due to the hidden variables, the representation power of the restricted Boltzmann machines 
is greater. It is proven that
with a sufficiently large number of hidden units any probability distribution is well approximated
\cite{LeRoux}, which is called a universal approximation theorem for the restricted Boltzmann machines. 
Adding more 
hidden layers can help the representation power
\footnote{Indeed, folding the deep Boltzmann machines leads to an RBM.}, and the following is called a deep Boltzmann machine \cite{dbm},
\begin{align}
{\cal E} \equiv 
\sum_{i,j} w_{ij}^{(0)} v_i h_j^{(1)} +
\sum_{k=1}^{N-1} 
\left[ 
\sum_{i,j} w_{ij}^{(k)} h_i^{(k)} h_j^{(k+1)} \right],
\label{dbmene}
\end{align}
where the index $k$ labels the hidden layers $k=1,2,3,\cdots, N$. 
The hidden variables $h_i^{(k)}$ in $N$ hidden layers, taking binary values, 
are summed as in \eqref{sumh}:
\begin{align}
P(v_i) = \sum_{h_i^{(k)}} \exp\left(-{\cal E}(v_i, h_i^{(k)})\right) \, .
\label{PDBM}
\end{align}
The visible layer consisting of
units whose values are the input $v_i$ may be thought of as $k=0$, 
which means $v_i = h_i^{(0)}$. 
The weights are again restricted as in the restricted Boltzmann machines, 
see Fig.~\ref{fig:bm} Right.

Although the number of units in each hidden layer may not be equal to each other,
in this paper we consider the case of having the same number of units in each layer,
for a structural simplicity.

%%%%%%%%%%%%%%%%%%%%%%

\section{AdS/CFT as a Boltzmann machine}
\label{sec:holo}

\subsection{Dictionary}

Let us first describe the similarity between the AdS/CFT correspondence
and the deep Boltzmann machine, and construct a dictionary
between the two schemes. 
In the AdS/CFT correspondence \cite{Maldacena:1997re},
the fundamental formula relating the boundary and the bulk is 
the GKP-W relation \cite{Gubser:1998bc,Witten:1998qj}
which is
\begin{align}
Z_{\rm QFT}[J] = \exp\left(-S_{\rm gravity}[\phi]\right) \, .
\label{AdScl}
\end{align}
This expression is for the large $N_c$ limit of the QFT with its generating functional $Z[J]$, 
while for the
finite $N_c$ this expression should be replaced by \footnote{We omit various details in 
this formula, such as the conformal dimension dependence.}
\begin{align}
Z_{\rm QFT}[J] = \int_{\phi(z=0)=J} \!\!\!\!\!{\cal D}\phi \, \exp\left(-S_{\rm gravity}[\phi]\right) \, .
\label{gkpwq}
\end{align}
Here $z$ is the emergent bulk coordinate, and $z=0$ is the boundary of the asymptotically 
AdS bulk, where the boundary condition $\phi(z=0)=J$ is put for the bulk field $\phi(x,z)$.

The deep Boltzmann machine approximates a given provability distribution 
by the formula \eqref{PDBM} with the energy function \eqref{dbmene}.
The similarity between the quantum gravity version of 
the GKP-W relation \eqref{gkpwq} and the definition equation of the deep Boltzmann machine
\label{PDBM} is obvious. The identification rules are as follows: 
the source function $J(x)$ is the input value $v_i$ of the visible layer, 
the bulk field $\phi(x,z)$ is the hidden variables $h_i^{(k)}$,
the emergent bulk coordinate $z$ is the label for the hidden layers $k$,
the generating function $Z[J]$ of the QFT is the provability distribution $P(v_i)$,
and the bulk action $S[\phi]$ is the energy function ${\cal E}(v_i, h_i^{(k)})$. See 
Table \ref{table:cor}
for a summary of the correspondence. The path integral of the bulk field $\phi$
is replaced by the summation over the hidden variables $h_i^{(k)}$,
so in general the quantum bulk of the AdS/CFT correspondence is a deep Boltzmann machine.

\begin{table}[t]
  \begin{tabular}{|c|c|}
      \hline 
      AdS/CFT & Deep Boltzmann machine \\
      \hline
Bulk coordinate $z$ & Hidden layer label $k$\\
      QFT source $J(x)$ & Input value $v_i$ \\
      Bulk field $\phi(x,z)$ & Hidden variables $h_i^{(k)}$ \\
QFT generating function $Z[J]$ & Provability distribution $P(v_i)$ \\
Bulk action $S[\phi]$ & Energy function ${\cal E}(v_i, h_i^{(k)})$\\
\hline
  \end{tabular}
  \label{table:cor}
\end{table}

The resemblance is basically the fact that the deep Boltzmann machine tries to reproduce
the probability distribution whose input is the values at the visible layer, and in the AdS/CFT
in the same manner, the bulk path-integration tries to reproduce
the generating functional of the boundary QFT where the input is the boundary value of the
bulk field.

In order to make the probability interpretation of the QFT generating functional, we normalize it
as
\begin{align}
P_{\rm QFT}[J]\equiv Z_{\rm QFT}[J]/Z_0\, , \quad Z_0 \equiv \int\!{\cal D}J\, Z_{\rm QFT}[J] \, .
\label{PQFT}
\end{align}
Then, using the deep Boltzmann machine representation of the bulk, training of the bulk theory
is possible to reduce the error function which is given by the Kullback-Leibler divergence
of the QFT partition function and the model probability of the Boltzmann machine,
\begin{align}
\mbox{Error fn.} = D_{\rm KL}\left(P_{\rm QFT}[J] \big|\big| P(v_i)\right) \, .
\label{KL}
\end{align}

As the deep Boltzmann machine allows arbitrary architecture for its neural network,
it is naturally expected that the AdS/CFT correspondence may be included as an example
of the Boltzmann machine. Below we shall demonstrate that a typical AdS/CFT model
allows a deep Boltzmann machine architecture.

%%%%%%%%%%%%%%%%%%%%%%%
\begin{figure*}[t]
\begin{center}
\includegraphics[width=10cm]{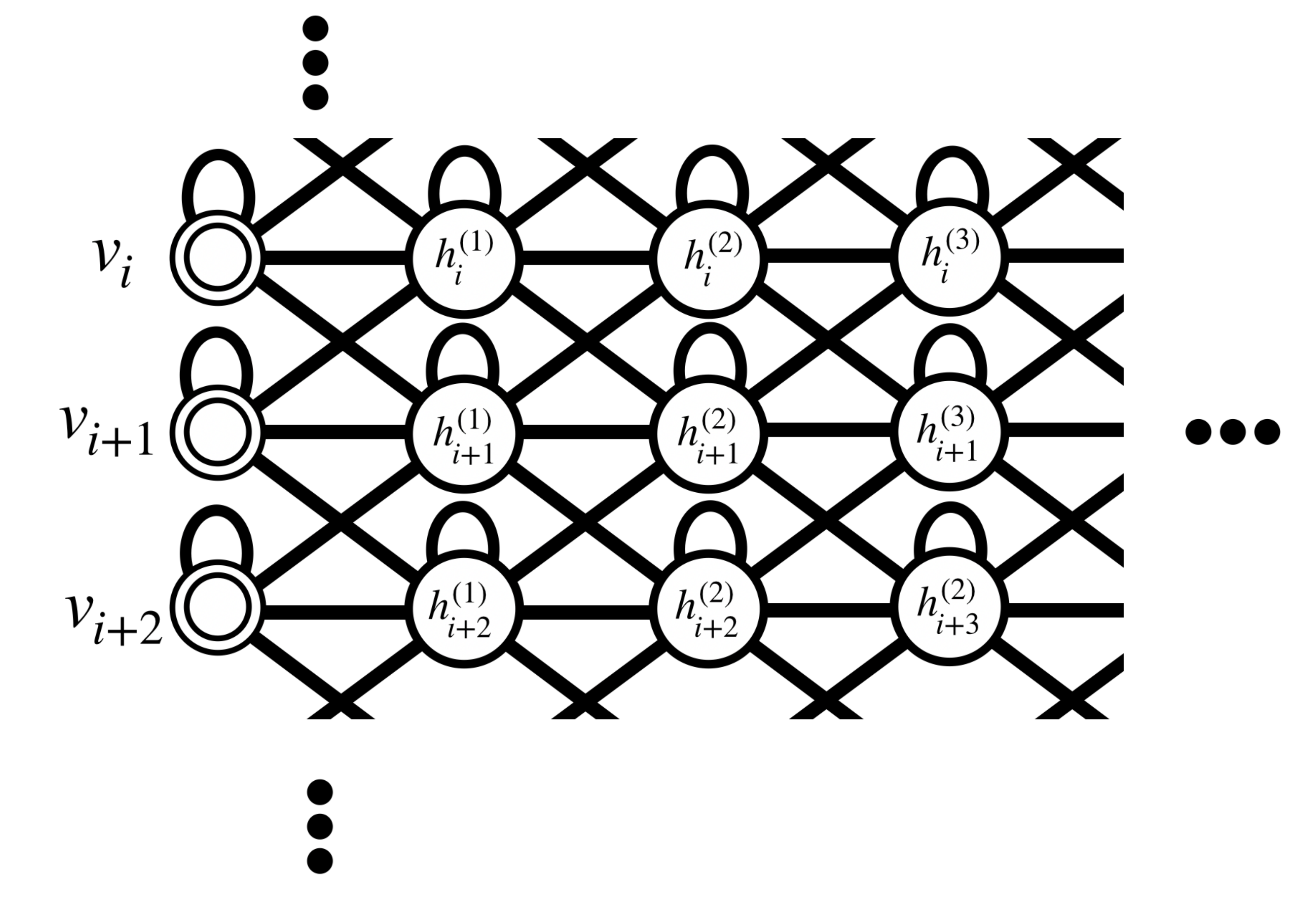}
\caption{
\label{fig:AdSBM}
The architecture of the deep Boltzmann machine for the AdS/CFT. The thick lines mean
weights. The difference from the standard Boltzmann machines in Fig.~\ref{fig:bm} is that
we allow a weight connecting the same unit, denoted as a curved line just above each unit
in the figure. We omit drawing the $l$-direction.}
\end{center}
\end{figure*}
%%%%%%%%%%%%%%%%%%%%%%%

%%%%%%%%%%%%

\subsection{Bulk as a neural network}
\label{subsec:bulk}

The simplest bulk action is for 
a free massive scalar field in an asymptotically AdS${}_{d+1}$ bulk geometry,
\begin{align}
S = & \int d^dx dz \, \frac12
\left[
 a(z) (\partial_z \phi)^2 + b(z) \sum_{I=1}^{d-1}(\partial_I \phi)^2 
\right.
\nonumber \\
& \left.
\hspace{20mm}
 + d(z) (\partial_\tau \phi)^2 + c(z) m^2 \phi^2
\right] \, .
\label{bulk-ac}
\end{align}
We chose Euclideanized signature so that the AdS/CFT correspondence can fit the scheme
of Boltzmann machines. The $d$-th coordinate $\tau \equiv x^d$ is the Euclideanized time coordinate.
Local interaction terms such as $\phi^n$ can be treated similarly below, but in this paper we
consider only the free case.

We assumed for simplicity that the metric depends only on the bulk emergent direction $z$ and is diagonal, and
assumed also a homogeneous spacetime  
about $x^I (I=1,2,\cdots, d-1)$, then $g_{11}(z) = \cdots = g_{d-1, d-1}(z)$. 
%With time-independent bulk scalar $\partial_{d}\phi=0$, 
We find
\begin{align}
& a(z) = [g_{11}(z)^{d-1} g_{dd}(z)/g_{zz}(z)]^{1/2} , \quad \\
& b(z) = [g_{11}(z)^{d-3}g_{dd}(z)g_{zz}(z)]^{1/2} , \quad \\
& c(z) = [g_{11}(z)^{d-1} g_{dd}(z)g_{zz}(z)]^{1/2}, \quad \\
& d(z) = [g_{11}(z)^{d-1} g_{zz}(z)/g_{dd}(z)]^{1/2} .
\end{align}
There exists a relation among them,
\begin{align}
(a(z)d(z))^2 = (c(z)/d(z))^{d-1} \, .
\label{relabcd}
\end{align}

In the standard Poincare coordinate system, the asymptotically AdS${}_{d+1}$
geometry is
\begin{align}
ds^2 = L^2 \frac{dz^2 + \sum_{\mu=1}^{d} (dx^\mu)^2}{z^2} \quad (z\sim 0)
\end{align}
with the AdS radius $L$, so we have the condition
\begin{align}
a\sim b\sim d\sim  (L/z)^{d-1}, \quad 
c\sim  (L/z)^{d+1}
\end{align}
near the AdS boundary $z\sim 0$.

Let us discretize the action \eqref{bulk-ac} to make it written like the energy function ${\cal E}$
of the deep Boltzmann machine.\footnote{
A continuum limit of the deep layers
was studied in a different context \cite{deepest}.}
First, the bulk geometry is discretized to a regular lattice whose sites are labeled by $(k,i,l)$;
The label $k$ refers to the discretized bulk emergent direction $z$, 
\begin{align}
z_k \equiv k \Delta z   \quad (k=0, 1,2,\cdots)
\end{align}
where $\Delta z$ is the lattice spacing.
In the same manner, we discretize $x^I$ and $\tau \equiv x^d$ by the lattice spacing 
$\Delta x$ and $\Delta \tau$, giving the label $i$ and $l$ respectively, as $x_{i,l}$.
This simplest regularization scheme replaces the integration over $d^dx dz$ by a sum $\sum_{k,i,l}$.

The bulk field $\phi(x,z)$ at the sites are written as
\begin{align}
h_{i,l}^{(k)} \equiv \phi(x_{i,l}, z_k). 
\end{align}
Thus the bulk scalar field is the variables in the hidden units.
Naturally, we identify the label $k$ as the label for the layers of a deep Boltzmann machine.
%where $x_i$ is the coordinate of the sites in the $x$ spacetime discretized with a lattice spacing $\Delta x$.
We define our visible layer as the AdS boundary value of the scalar field, {\it i.e.} the first $k=0$ component of $h$,
\begin{align}
v_{i,l} \equiv h_{i,l}^{(0)} \, . 
\label{vh}
\end{align}

The $z$-derivative term in the bulk Lagrangian is replaced by
\begin{align}
(\partial_z \phi)^2 = \lim_{\Delta z \to 0} \frac{\left(\phi(z_{k+1})-\phi(z_k)\right)^2}{(\Delta z)^2} \, .
\end{align}
As for the derivative terms concerning $\partial_\tau$ (and similarly for $\partial_I$), we choose
\begin{align}
& (\partial_\tau \phi)^2 =\lim_{\Delta \tau,\Delta z \to 0} 
\left[
\frac{\phi(x_{i,l+1},z_k)-\phi(x_{i,l},z_k)}{\Delta \tau}
\right.
\nonumber \\
& \hspace{25mm} 
\left.
\cdot \frac{\phi(x_{i,l+1},z_{k+1})-\phi(x_{i,l},z_{k+1})}{\Delta \tau}
\right] .
\label{chose}
\end{align}
Note the dependence on the label $k$; the reason we chose this discretization will be clear below.

The background metric functions are discretized in the same manner,
\begin{align}
& a_k \equiv a(z=z_k), \quad 
b_k \equiv b(z=z_k) , \nonumber \\
& c_k \equiv c(z=z_k) \quad 
d_k \equiv d(z=z_k).
\end{align}
Then the bulk action is written as
\begin{align}
S = &
\sum_{k,i,l} 
\left[
a_k \frac{1}{2(\Delta z)^2} \left(
h_{i,l}^{(k+1)}-h_{i,l}^{(k)}
\right)^2
+ c_k \frac{m^2}{2} \left(h_{i,l}^{(k)}\right)^2
\right.
\nonumber 
\\
& + 
b_k \frac{1}{2(\Delta x)^2} 
\left(h_{i+1,l}^{(k)}-h_{i,l}^{(k)}\right)
\left(h_{i+1,l}^{(k+1)}-h_{i,l}^{(k+1)}\right)
\nonumber 
\\
& + 
\left.
d_k \frac{1}{2(\Delta \tau)^2} 
\left(h_{i,l+1}^{(k)}-h_{i,l}^{(k)}\right)
\left(h_{i,l+1}^{(k+1)}-h_{i,l}^{(k+1)}\right)
  \right] .
\end{align}
This is recast to the following Boltzmann machine form:
\begin{align}
S = {\cal E} \equiv \sum_k 
\left[ 
\sum_{i,j}\sum_{l,m} 
\left\{
w_{ij,lm}^{(k)} h_{i,l}^{(k)} h_{j,m}^{(k+1)} 
\right.\right.
\nonumber
\\
\left.\left.
\hspace*{20mm}
+ 
\widetilde{w}_{ij,lm}^{(k)}h_{i,l}^{(k)} h_{j,m}^{(k)} 
\right\}
\right], 
\label{bulkww}
\end{align}
where the weights are given as
\begin{align}
w_{ij,lm}^{(k)} \equiv 
&
-\frac{a_k}{(\Delta z)^2}\delta_i^j \delta_l^m
\nonumber 
\\
&
+\frac{b_k}{2(\Delta x)^2}
\left(
2\delta_i^j\delta_l^m-\delta_{i+1}^j\delta_l^m-\delta_{i}^{j+1}\delta_l^m
\right)
\nonumber \\
&+\frac{d_k}{2(\Delta \tau)^2}
\left(
2\delta_i^j\delta_l^m-\delta_i^j\delta_{l+1}^m-\delta_i^j\delta_l^{m+1}
\right)
, 
\label{w}
\\
\widetilde{w}_{ij,lm}^{(k)} \equiv 
&
\left( \frac{a_k\!+\!a_{k-1}}{2(\Delta z)^2} +m^2\frac{c_k}{2}\right) 
\delta_i^j\delta_l^m \, .
\label{wtil}
\end{align}
These weights are symmetric.

The path integral over the bulk field $\phi(x^I,\tau,z)$ is equivalent to the integration over
all the hidden variables $h_{i,l}^{(k)}$ $(k=1,2,\cdots)$, therefore
the GKP-W relation \eqref{gkpwq} is written as 
\begin{align}
Z_{\rm QFT}[J] = \sum_{h} \, \exp\left(-{\cal E}\right)
\end{align}
where ${\cal E}$ is defined by \eqref{bulkww}. 
And through \eqref{vh} we have
\begin{align}
J(x_{i,l}) = v_{i,l} = h_{i,l}^{(0)}.
\end{align}
This is a deep Boltzmann machine representation of 
the AdS/CFT correspondence.
See Fig.~\ref{fig:AdSBM} for our architecture.

The background metric appears as the weights of the Boltzmann machine.
As is understood from \eqref{w} and \eqref{wtil}, the weights are
not all independent. They form quite a sparse neural network. The trained variables are 
$a(z_k)$, $b(z_k)$, $c(z_k)$ and $d(z_k)$ $(k=0,1,2,\cdots)$, under the constraint \eqref{relabcd}.
The bulk scalar field appears as the hidden variables to be summed,
at which the boundary value of the bulk scalar field is identified with
the visible units. 

Note that,
because we chose the discretization scheme \eqref{chose},
 the weights $\tilde{w}$ connecting the units in the same layer are completely diagonal
 (of the form $\delta_i^j \delta_l^m$) as seen in \eqref{wtil}. As explained earlier, this
does not violate the conditional independence of the units in the same layer,
which is important for the training of the Boltzmann machine.

\subsection{Discretized values of the bulk field}
\label{sec:disc}

Standard Boltzmann machines allow binary values for the variables $h$,
while the AdS/CFT correspondence requires continuous values for the bulk field $\phi(x,z)$. 
To bridge these two, we need to discretize also the field value space.
Suppose that typical values necessary for training the Boltzmann machines
are in the range $|\phi| < A$. Then a natural discretization of the values
is given as
\begin{align}
\phi = A \frac{u}{u_0}
\end{align}
where $u= -u_0, -u_0+1, \cdots, 0, \cdots, u_0-1$. In this discretization, we have 
$2u_0$ different values for $\phi$ to take. To bring them to a set of binary-valued
variables, we introduce the binary variable $s_{(\tilde{u})} \in \{0,1\}$ as
\begin{align}
\phi = A \left(
-1 + \frac{2\tilde{u}}{u_0} + \frac{1}{u_0} s_{(\tilde{u})}
\right) \, .
\end{align}
Here we divided the single entry $\phi$ to $u_0$ entries $s_{(\tilde{u})}$
with $\tilde{u}=0,1,\cdots,u_0-1$. In effect, each unit referring to $h_{i,l}^{(k)}$
in the Boltzmann machine is split into $u_0$ different units $s$.
All of those split units need to share the same weight for every original connection
with different unit $h$. In this manner, binary-valued Boltzmann machines can be
constructed from the continuous-valued Boltzmann machines.

%%%%%%%%%%%%%%

\subsection{The deepest layer is the end of space}

In the AdS/CFT correspondence, the IR end of the geometry is important, as
it directly reflects the properties of allowed spectra of the QFT. Popular 
holographic geometries are confining geometries and black holes,
and they have specific boundary conditions at the IR end of the geometry.
Except for the cases of conformal field theories as the boundary QFT,
the bulk geometry naturally terminates at some IR scale $z=z_{\rm IR}$.
In the terminology of the deep Boltzmann machines, this means that
the layers terminate at $k=N$ with $N\equiv z_{\rm IR}/\Delta z$.
Let us rephrase those geometric boundary conditions to the
treatment around the deepest layer $k=N$ of the deep Boltzmann machine.

First of all, the layers actually
terminates at $k=N$, and there is no additional layer at $k=N+1$.
In terms of the weights, this condition means $w_{ij,lm}^{(N)}=0$, which is
\begin{align}
a(z_N)=b(z_N)=d(z_N)=0 \, .
\end{align}

The confining geometry refers to the Dirichlet boundary condition
for the bulk field $\phi$, as it simply means that 
the bulk field $\phi$ needs to vanish in %there does not exist 
the spacetime in the region specified by $z>z_{\rm IR}$. This
location is called a ``hard wall" in holography. 
%and a popular example is
%the geometry provided in \cite{Witten:1998zw}.
%The IR bulk Dirichlet boundary condition means that 
%
In general, the condition of the hard wall means that the metric function
which the scalar field feels has a special behavior there.
In fact, to impose $\phi(z_{\rm IR})=0$ we just need that the mass
$c(z) m^2$ at $z=z_{\rm IR}$ diverges. So, in this case 
we can rephrase the Dirichlet boundary condition in terms of the metric function:
\begin{align}
c(z_{N}) = \infty \, .
\label{bccon}
\end{align}

%
%\begin{align}
%h_i^{(k=N)} = 0 \quad (\mbox{for any } i)
%\label{bccon}
%\end{align}
%where $N \Delta z = z_{\rm IR}$ is the IR end of the geometry.
%This condition, $\phi(z=z_{\rm IR})=0$, naturally comes from
%the finiteness of the bulk action under the IR boundary behavior
%of the bulk metric,
%\begin{align}
%b(z=z_{\rm IR}) =c(z=z_{\rm IR}) = \infty.
%\end{align}
%Note that $a(z=z_{\rm IR})$ needs not to diverge, and therefore,
%there is no constraint on $\tilde{h}^{(N)}$ in the Boltzmann machine.
%See Fig.~\ref{fig:deepest} for this deepest layer architecture.
%Of course, we need $h^{(N+1)}=\tilde{h}^{(N)}=0$ since at $k=N+1$
%there is no spacetime.
%
%
%
%%%%%%%%%%%%%%%%%%%%%%%%
%\begin{figure}[t]
%\begin{center}
%\includegraphics[width=6.5cm]{res.pdf}
%\caption{
%\label{fig:deepest}
%Top: the AdS/CFT correspondence. Bottom: a deep Boltzmann machine. 
%}
%\end{center}
%\end{figure}
%%%%%%%%%%%%%%%%%%%%%%%%

Next, consider the black hole horizon condition instead.
At the black hole horizon the $zz$ component of the metric diverges,
while the temporal $dd$ component vanishes. 
Thus $a(z)=0$, and $d(z)$ diverges, while $b(z)$ and $c(z)$, and $a(z)d(z)$ remain finite
and nonzero. Therefore, the black hole boundary condition is
\begin{align}
a(z_{N-1})\propto \Delta z, \quad d(z_{N-1})\propto \frac{1}{\Delta z} \, ,
\label{bcbh}
\end{align}
with infinitesimally small $\Delta z$.

The confining condition \eqref{bccon} and
the horizon condition \eqref{bcbh} are examples of more general constraints.
We can impose other boundary conditions if they are consistent with
the large $N_c$ limit of the AdS/CFT, as we shall study in the next section.

For the pure AdS geometry, there is no IR end of the
space, and the $z$ direction is extended to $z=\infty$. So, to host all
possible asymptotically AdS spacetimes in our Boltzmann machine architecture, 
we need to prepare infinitely deep Boltzmann machines.\footnote{
However, note that $z=\infty$ corresponds to the strictly vanishing energy in the QFT,
which are not the main constituent of the partition function $Z[J]$ except for
IR regularization ambiguities.}

%%%%%%%%%%%%%%%%%%%%%%%
\begin{figure*}[t]
\begin{center}
\includegraphics[width=12cm]{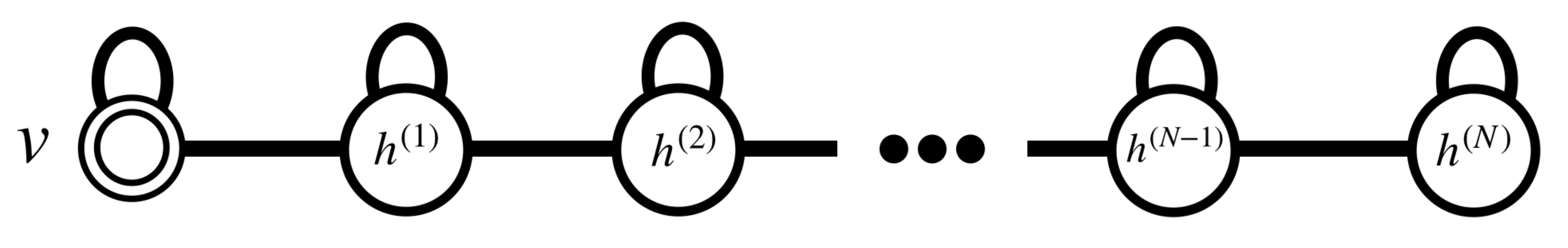}
\caption{
\label{fig:onshell}
The simplified deep Boltzmann machine at which the $i$- and $l$-dependence 
are ignored.
}
\end{center}
\end{figure*}
%%%%%%%%%%%%%%%%%%%%%%%

%%%%%%%%%%%%%%

\section{Saddle point of Boltzmann machine}
\label{sec:saddle}

The AdS/CFT correspondence has been studied in the large $N_c$ limit of the QFT,
because it is the classical limit of the bulk which is the only reliable gravity calculation, 
in the absence of satisfactory quantum gravity formulation. The large $N_c$ limit, or the classical limit
of the gravity theories, 
is equivalent to the zero temperature limit of the Boltzmann machine, ${\cal E}$
replaced by ${\cal E}/T$ and $T \to 0$.
At the limit, gravity theory can be well approximated by saddle points --- the
solutions of the classical equations of motion, and the on-shell action
is simply substituted to the right hand side of \eqref{AdScl}.

The zero-temperature limit of Boltzmann machines has not been studied extensively,
because the hidden/visible variables in ordinary Boltzmann machines 
take only binary values and the saddle approximation is not effective. 
In our case, as described in Sec.~\ref{sec:disc}, we consider a certain limit
of binary-valued Boltzmann machines to acquire continuous-valued variables. There
the equations of motion, and the saddle points, make sense. In this section, we study 
consistency conditions of the classical limit (equivalently, the zero temperature limit, or the saddle point approximation) 
of the 
deep Boltzmann machine given in the previous section. 
For simplicity we treat the variables $h$ as continuous variables.
 
First, let us consider the standard restricted Boltzmann machine \eqref{rbm} with continuous-valued 
variables, and 
how the classical limit causes an inconsistency.  The saddle point equation is
\begin{align}
0 = \frac{\delta {\cal E}}{\delta h_i}
= b_i + \sum_{j} w_{ji} v_j \, .
\end{align}
Since $b_i$ and $w_{ji}$ are the parameters to be fixed after the training with 
various sets of $\{v_i\}$,
this equation cannot be satisfied. Therefore, restricted Boltzmann machines
with continuous hidden variables do not allow the saddle point approximation,
on the contrary to the physical intuition.

Adding more hidden layers can resolve the issue. Suppose we have another hidden layer
to the restricted Boltzmann machine, 
\begin{align}
{\cal E} \equiv 
& \sum_{i} (a_i v_i + b_i h_i^{(1)} + c_i h_i^{(2)})
\nonumber \\
&+ 
\sum_{ij} 
\left(w_{ij}^{(0)} v_i h_j^{(1)}  +w_{ij}^{(1)} h_i^{(1)} h_j^{(2)} \right) \, .
\label{rrbm}
\end{align}
Then the saddle point equation is
\begin{align}
& 0 = \frac{\delta {\cal E}}{\delta h_i^{(1)}} = 
b_i + \sum_{j} 
\left(w_{ji}^{(0)} v_j  +w_{ij}^{(1)} h_j^{(2)} \right) \, , 
\\
&0 = \frac{\delta {\cal E}}{\delta h_i^{(2)}} = 
c_i + \sum_{j} 
w_{ji}^{(1)} h_j^{(1)}  \, .
\end{align}
The first equation determines $h_j^{(2)}$ for any given training value of $v_i$,
so it gives a consistent saddle point equation. The second equation simply shows 
that the middle layer variable $h^{(1)}$ takes a fixed value $- [[w^{(1)}]^T]^{-1} c$.
So, substituting these to the original energy function \eqref{rrbm}, we obtain 
the saddle point approximation of the restricted Boltzmann machine,
\begin{align}
& {\cal E}_{\rm on-shell} = {\cal E}_0+ 
\sum_{i} a^{\rm eff}_{i}v_i   \, , 
\\
&{\cal E}_0 \equiv  -\sum_{ij}(w^{(1)})^{-1}_{ij}c_i b_j \, ,
\\
&
a^{\rm eff}_{i} \equiv a_i - \sum_{j} c_j \left(
(w^{(1)})^{-1}(w^{(0)})^T
\right)_{ji} \, .
\end{align}
Here it should be noted that the obtained energy function is linear in $v$, 
so it does not have the form of the standard Boltzmann machines
whose energy functions are bilinear in $v$.
The reason of the linearity is that the saddle point equations for the $k$-odd
and the $k$-even layers decouple from each other.

Instead of adding more layers, we can introduce a self-coupling $\delta_i^j h_i h_j$
as described earlier in Sec.~\ref{sec:review}.
For the case with just a single hidden layer with a uniform self-coupling weight,
we have
\begin{align}
{\cal E} \equiv \sum_i \left(
a_i v_i + b_i h_i
\right) + \sum_{ij} 
\left(w_{ij}v_i h_j + c \, \delta_{ij}h_i h_j\right)\, .
\label{Ehh}
\end{align}
The saddle point equation is
\begin{align}
0 = \frac{\delta {\cal E}}{\delta h_i} = 
b_i + 2 c h_i + \sum_{j} 
\left(w_{ji}v_j   \right) \, , 
\end{align}
which determines the value of the hidden unit $h_i$ in terms of the input $v_i$,
so it gives a consistent solution. The on-shell value of the energy function is
\begin{align}
{\cal E}_{\rm on-shell} = {\cal E}_0+ 
\sum_{ij} a^{\rm eff}_{i}v_i 
+ 
\sum_{ij} w^{\rm eff}_{ij}v_iv_j   \, , 
\end{align}
where
\begin{align}
&{\cal E}_0 \equiv  -\frac{1}{4c} \sum_{i}b_i b_i \, ,
\\
&
a^{\rm eff}_{i} \equiv a_i - \frac{1}{2c}\sum_{j} w_{ij}b_j 
 \, ,
\\
&
w^{\rm eff}_{ij} \equiv -\frac{1}{4c} \left(w w^T\right)_{ij} \, .
\end{align}
Thus, as is expected, the effective energy function is bilinear in $v_i$. 

Keeping these results in mind, we consider the deep Boltzmann machine
which we defined in the previous section.
The saddle point condition is
\begin{align}
 0 = & \frac{{\cal E}}{\delta h_{i,l}^{(k)} }
 \nonumber \\
= & \sum_{j,m}
\left[ 
w_{ji,ml}^{(k-1)} h_{j,m}^{(k-1)}  + 
w_{ij,lm}^{(k)} h_{j,m}^{(k+1)} +\tilde{w}^{(k)}_{ij,lm}h_{j,m}^{(k)} \right] \, .
\label{heq}
\end{align}
So, the variables at the layer $k$ are related to those of the layer $k+1$ 
and of the layer $k-1$. The equation has both the properties of the cases of
\eqref{rrbm} and \eqref{Ehh}.

Let us study the consistency with the IR boundary condition, the deepest layer.
For simplicity, to look at the consistency, we consider the case with a homogeneous $\phi$
in $x^I$ and $x^d$, which is equivalent to ignore the terms with $b(z)$ and those with $d(z)$.
The architecture of the deep Boltzmann machine is shown in Fig.~\ref{fig:onshell}.
At the deepest layer $k=N$, the saddle point equation gives
\begin{align}
h^{(N-1)} \sim  h^{(N)}
\end{align}
where the symbol $\sim$ denotes a linear relation whose coefficients are given by weights.
Similarly, using the saddle point equation at $k=N-1$ which gives
\begin{align}
h^{(N-1)} + h^{(N-2)} \sim  h^{(N)}
\label{N-2}
\end{align}
where we omit the coefficients. Then altogether, they give $h^{(N-2)} \sim  h^{(N-1)}$.
Repeating this backwards in layers, we finally obtain
\begin{align}
h^{(0)} \sim  h^{(1)} 
\end{align}
which can also be written as
\begin{align}
v \sim  \frac{h^{(1)}-h^{(0)} }{\Delta z} \, .
\end{align}
In the continuum limit, this relation is
\begin{align}
\phi(z=0) \sim  \partial_z \phi(z=0). 
\end{align}
In the boundary QFT of the AdS/CFT correspondence, 
this relation is equivalent to the linear response relation
\cite{Klebanov:1999tb},
\begin{align}
J \sim \langle {\cal O}\rangle \, .
\end{align}
Thus, the deep Boltzmann machine is found to be consistent with the standard analysis
in the classical bulk side of the AdS/CFT correspondence.

It is intriguing that the saddle point approximation provides explicitly the relation between 
the variables at the adjacent layers. This relation is expected for neural networks of the 
feed-forward type. So, we find that the saddle point approximation of the deep Boltzmann machine
provides a feed-forward architecture.
A subtle difference from the standard feed-forward is that the linear relation starts at
the deepest layer, not at the visible layer. In fact, looking at only the first hidden layer we find that 
the relation is just 
like \eqref{N-2}, so it is not a linear relation between just the adjacent two layers. In fact,
the scalar field equation is the second order differential equation, so, there is a backward wave
in addition to the forward wave. These two waves satisfy the consistency condition at the deepest
layer. Therefore, the saddle point approximation provides a ``folded feed-forward" structure.
Unfolding the folded structure is possible, and in Appendix \ref{sec:Auto} we provide an architecture
of the unfolded type, which looks like an autoencoder.

%%%%%%%%%%%%%%

\section{Regularization and Einstein action}
\label{sec:Ein}

In this section we study the condition for the trained weights to be interpreted as a bulk spacetime.
The training should be performed in the following manner. First, prepare a quantum field theory for
which one wants to know whether a gravity dual exists or not. Then calculate $Z_{\rm QFT}[J]$ and
its probability interpretation $P_{\rm QFT}[J]$ by \eqref{PQFT} \footnote{For this, one may need 
to use some approximation and some assumption on superselection sectors to perform the path integrations. }.
Prepare the deep Boltzmann architecture given in Sec.~\ref{subsec:bulk}, 
and by updating the weights to reduce the KL divergence \eqref{KL}. Once the KL divergence decreases
to enough accuracy, we say that the bulk is learned.

The metric function is encoded in the sparse weights $w, \tilde{w}$  in the deep Boltzmann machine, given in
\eqref{w} and \eqref{wtil}. Although it can be
easily reconstructed, there is one issue: generically, the training ends up with various different sets of weights,
because the error function may have many almost degenerate local minima. Each of the local minima can
approximate $P_{\rm QFT}$ very well --- which is related to the notion ``generalization" in machine learning.
%In this sense, each deep Boltzmann machine is an alternative
%representation of the QFT.

%However, 
We are looking for a gravity dual. For the trained Boltzmann machine weights to be interpreted as
a bulk spacetime, we need a criterion to pick up a certain set of the weights among the degenerate 
local minima. The criterion is simple: use an Einstein action for a regularization of the deep Boltzmann machine.\footnote{
The word ``regularization" here is for Tikhonov regularization, not for the lattice regularization.}

Basically, the generic trained weights take quite scattered values, and they are not a smooth function of $z$
in the continuum limit $\Delta z \to 0$. 
For those configurations of weights, the Einstein action takes a large value. On the other hand, smooth
metric functions of $z$, and so a set of the weights whose values do not drastically vary as one sweeps the depth of the layers, 
have lower values of the Einstein action. Therefore, the Einstein action can be used for 
selecting a proper set of weights which has a bulk spacetime interpretation.

A proposed regularization term is a discretization of the Einstein action with a negative cosmological term,
\begin{align}
{\cal E}_{\rm reg} = \int \! d^{d}xdz \sqrt{\det g}\left(R +\frac{d(d-1)}{ L^2}\right) \, .
\end{align}
To obtain the explicit discretization, for simplicity we consider a conformal spacetime
\begin{align}
g_{\mu\mu}=g_{zz} = \frac{L^2}{z^2}(1 + \alpha(z)) \, .
\end{align}
When $\alpha(z)=0$, the metric reduces to the pure AdS metric. So the asymptotically AdS
spacetimes allow only $\lim_{z\to 0}\alpha(z) = 0$.
In terms of the previous $a(z)$, $b(z)$, $c(z)$  and $d(z)$, this ansatz leads to
\begin{align}
&a= b=d= (L/z)^{d-1} (1+\alpha(z))^{(d-1)/2}, \nonumber \\
&c=  (L/z)^{d+1}(1+\alpha(z))^{(d+1)/2} \, .
\end{align}
So, the discretization of the $z$ direction as $z=z_k$ provides
a lattice on which $\alpha_k \equiv \alpha(z_k)$ is defined for $k=0,1,2,\cdots$.
For $d=3$ as an example, 
the Einstein action becomes
\begin{align}
\frac{{\cal E}_{\rm reg}}{3L^2}
=\int \! d^{d}xdz \left[
\frac{-2}{z^4} + \frac{2}{z^2}\alpha(z)^2 + \frac{1}{2z^2}\frac{\left(\alpha'(z)\right)^2}{1+\alpha(z)}
\right]. 
\label{Ereg1}
\end{align}
Discretizing this action, we obtain the regularization term is the error function
\begin{align}
{\cal E}_{\rm reg}
= c \sum_{k=1}^\infty
\left[
\frac{2}{k^2}\alpha_k^2 + \frac{1}{2k^2}\frac{\left(\alpha_{k+1}-\alpha_k\right)^2}{1+\alpha_k}
\right].
\label{Ereg2}
\end{align}
Here $c$ is a positive constant, and we ignored an additive constant term which is
irrelevant for the training. 
The additive constant comes from the first term in \eqref{Ereg1},
that is, the cosmological constant for the pure AdS spacetime.

%We call this regularization \eqref{Ereg2} as Einstein regularization. 
It is easy to see that this regularization in fact favors a smoother distribution of the weights,
due to the second term in \eqref{Ereg2}. Using this regularization, during the training,
the Boltzmann machine tries to minimize also the Einstein action at the same time. When the
error decreases to a satisfactory small value, the weights can be interpreted as an Einstein
spacetime.\footnote{Note also that the black hole horizon behavior of the weights, \eqref{bcbh},
is consistent with this regularization.}

When ${\cal E}_{\rm reg}=0$, we have a pure AdS geometry. In generic AdS/CFT correspondence,
the bulk action can take various forms; it may have more supergravity fields, and it may suffer from
higher derivative terms coming from quantum gravity corrections or stringy corrections.
Therefore, in general, the regularization needs to allow more generic actions, such as
\begin{align}
{\cal E}_{\rm reg} = \int \! d^{d}xdz \sqrt{\det g}\left(c_0 + c_1 R + c_2 R^2 + c_3 R^3 + \cdots \right) \, ,
\end{align}
or more with tensorial structures. This can be discretized by the same method, and we obtain
a more general Einstein regularization. Note that here in the expression the coefficients
$c_i$ are trained variables. In general we do not know the bulk gravity action, so we need to allow general 
action. When we say that the bulk is a spacetime, it means that it reduces the value of this
general action. When the powers of the Riemann tensors stop at some fixed value,
a low energy effective spacetime interpretation is possible.

%%%%%%%%%%%%%%%

\section{Summary and discussions}
\label{sec:Sum}

In this paper, we have shown that the standard AdS/CFT correspondence can be regarded
as a deep Boltzmann machine. The neural network architecture, once properly defined,
is interpreted as a bulk spacetime geometry. The network depth is the emergent direction in the bulk,
and the network weights are metric components.
Hidden variables correspond to discretized fields in the bulk, and the probability distribution
given by the Boltzmann machine is the generating functional of the QFT dual to the bulk gravity.

For the mapping we used a bulk scalar field theory in curved geometries. The IR boundary conditions
of the bulk, such as the black hole horizon or the hard wall, can be implemented to the weight behavior around
the deepest layer of the Boltzmann machine. The large $N_c$ limit of the AdS/CFT is argued in
the scheme of the Boltzmann machine, consistently giving an organized set of linear equations among weights.

Among many degenerate vacua of the deep Boltzmann machine, a set of weights which allows 
a spacetime interpretation is selected by a regularization in the error function in addition to the KL divergence. 
We have introduced  a natural regularization based on the Einstein action and its generalization.

Our study provides a relation between the AdS/CFT correspondence and the deep Boltzmann machine.
In view of the history of the quantum gravity, introducing discretization of the spacetime is natural, and
we hope that more concepts on Boltzmann machines and deep learning can be imported to quantum gravity, so
that it may shed light on the mystery of the bulk emergence in the holographic principle.

Several clarification and comments are in order. First, the discretization of the spacetime used in
this paper favors a certain coordinate system, and thus the general coordinate transformation
of the gravity theory is not seen in our framework. Furthermore, even the isometry transformation,
which is the scale transformation in the QFT, is difficult to be implemented in our formulation.
A hyperbolic network (as used in \cite{You:2017guh}) is better to be consistent with the isometry, but is difficult to
find a continuum limit. It would be interesting to seek for a more desirable discretization scheme.
In fact, a well-known approach for quantum gravity uses dynamical triangulation 
\cite{Regge:1961px} in which
connection bond topology (which is axon topology in neural networks) 
is dynamical. On the other hand
standard neural networks have a fixed architecture while the weights are variable. We may need
a refined discretization architecture, to have a more unified view of the quantum gravity and the deep learning.
Generic quantum gravity may include even non-geometric landscape for which machine learning have been applied
\cite{He:2017aed,He:2017dia,Liu:2017dzi,Carifio:2017bov,Ruehle:2017mzq,Faraggi:2017cnh,Carifio:2017nyb},
and a possible relation to our approach of having neural network as a spacetime would be interesting.

Second, we have introduced the saddle point approximation in the evaluation of the deep Boltzmann machine,
based on the standard large $N_c$ argument of the AdS/CFT correspondence.
At the limit, linear relations among weights at layers close to each other are derived,
and the information at the visible layer is processed through the bulk, as if it propagates.
This means that the saddle point of the deep Boltzmann machine brings it to a folded feed-forward type
deep neural network. The AdS/CFT interpretation of a feed-forward neural network was studied in 
\cite{Hashimoto:2018ftp,Hashimoto:2018bnb} and the trained
weights exhibit an interesting physical picture.

On the other hand, at finite $N_c$ (beyond the classical limit of the bulk), the relation between the
AdS/CFT and the deep Boltzmann machine is a little ambiguous --- in the bulk, only the scalar
field ($\sim$ hidden variables $h$) is path-integrated while the metric ($\sim$ weights $w$) is not.
This situation can be interpreted as that the scalar field is that of a probe brane in the bulk.
What is the metric path-integral in the deep Boltzmann machine? It is a statistical summation
of the network weights, which has been studied as statistical neural networks \cite{Amari1}. 
It would be interesting to see more connection between the holographic principle and the statistical
neural networks. In fact, in \cite{Amari2} a conformal transformation of data space was found at 
a layer-to-layer propagation, and it may allow a holographic interpretation.

Finally, we make a comment on a relation to quantum information. It is known that 
the AdS/CFT correspondence has a close relation to quantum information, in particular AdS/CFT
toy models based on tensor networks have been studied. The structure of MERA \cite{Swingle:2009bg} 
has a bulk hyperbolic space interpretation, and tensor networks using perfect tensors 
\cite{Pastawski:2015qua} provide
a quantum correspondence between the bulk and the boundary in the AdS/CFT. Since it is known \cite{Gao}
that
any quantum code allows its deep Boltzmann machine interpretation, the AdS tensor networks
can be mapped to deep Boltzmann machines. In general obtained machine architecture tends to
be complicated (since the number of quantum gates necessary to reproduce $A$-leg tensor 
is $\sim {\cal O}(2^{2A})$), so a continuum limit to have a continuum field theory in the bulk,
which we studied in this paper, is difficult to take. Further studies for bridging the holographic principle
and deep learning are desired.

\vspace{5mm}
\noindent
Note added: while this manuscript was prepared, we noticed that Ref.~\cite{Hu:2019nea} 
which interprets the bulk as a deep generative model was submitted to arXiv recently.

%%%%%%%%%%%%%%%%%%%%%%%%%
%\acknowledgments
\begin{acknowledgments}
The author would like to thank Shun-ichi Amari, Masatoshi Imada, 
Akinori Tanaka, Akio Tomiya and Yi-Zhuang You for valuable discussions.
The work of K.H. was supported 
in part by MEXT/JSPS KAKENHI Grants No.~JP15H03658, No.~JP15K13483 and No.~JP17H06462.
\end{acknowledgments}

\vspace{5mm}

%%%%%%%%%%%%%%%%%%%%%%%%%%%%%%%

\appendix

\section{Holographic autoencoder}
\label{sec:Auto}

In this appendix 
we describe an implementation of the AdS/CFT correspondence
into a feed-forward deep neural network of an autoencoder-like architecture. 
We discuss only the classical limit of the bulk (the large $N_c$ limit)
to make sure that the feed-forward structure is clear in the AdS/CFT.

\begin{figure*}
\includegraphics[width=12cm]{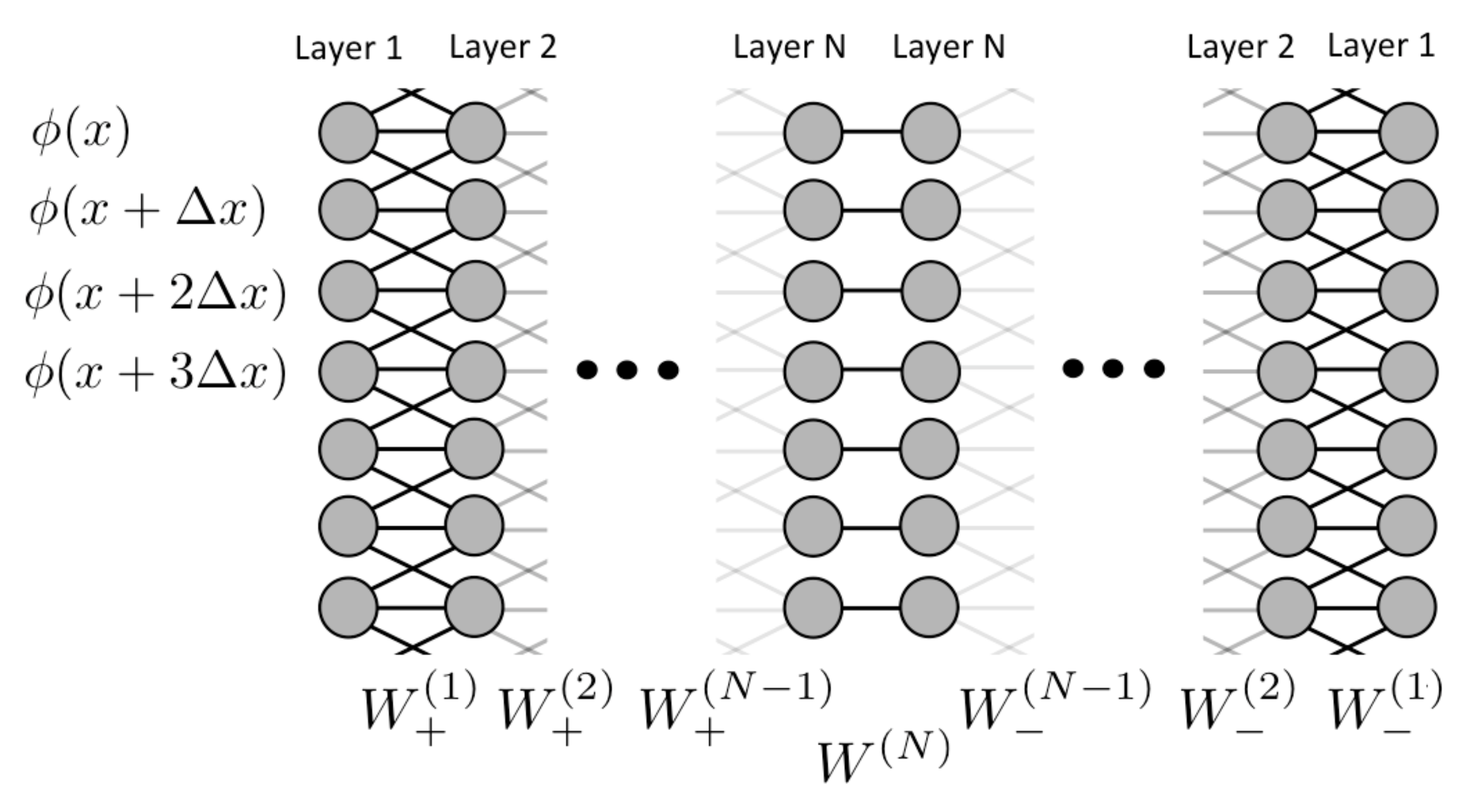}
\caption{Holographic autoencoder. The depth of the lines show average weights, which decrease
toward the black hole horizon (the neck part of the neural network).}
\label{fig:ae}
\end{figure*}

In \cite{Hashimoto:2018ftp,Hashimoto:2018bnb}, 
a deep neural network employs $J$ and $\langle{\cal O}\rangle$ as
the input at the initial layer while the black hole horizon boundary condition is used as an output at the final layer.
Another natural implementation is to use $J$ (the non-normalizable mode of the AdS scalar field) as
an input 
and $\langle{\cal O}\rangle$  (normalizable mode) as the output data.
In this case the data propagates from the AdS boundary toward the black hole horizon first, then it bounces
with the boundary condition, then propagates back to the AdS boundary again. Therefore the neural network
is an hourglass-type, and is naturally interpreted as an autoencoder in machine learning, see Fig.~\ref{fig:ae}.
The neural network looks similar to the two-sided black hole geometry which is often used in finite temperature
holography \cite{Maldacena:2001kr, Hartman:2013qma, Matsueda:2012xm, Mollabashi:2013lya}.

%%%%%%%%%%%%%%%%%%%%%%%
%%%%%%%%%%%%%%%%%%%%%%%

The important point for the construction is to use
the holographic renormalization group 
\cite{deHaro:2000vlm}
%,Skenderis:2002wp,Papadimitriou:2004ap,Papadimitriou:2004rz,Papadimitriou:2010as} 
to divide the second-order differential equation in $z$ to a set of two first-order equations
(for the non-normalizable and normalizable modes).
Generic autoencoders have two important features: their weights are left-right symmetric, and
they reduce dimensions of the data space at the neck of the network. 
In our holographic autoencoder, the left and right are governed by the same metric, 
and the convolution near the neck is red-shifted so that the weights effectively reduce at the neck.
%The details of the holographic autoencoder is provided in the supplemental material.

%%%%%%%%%%%%%%%%%%%%%%%%%%%%%%%%%%%%%%
%\subsection{Deep neural network for the holographic autoencoder}

Here we present details of a construction of the holographic autoencoder.
First, we explain how the second-order differential equation of the scalar field in the bulk can be 
equivalently replaced by a set of two first-order differential equations. 
This is important for the implementation of the
scalar system in the form of a deep neural network, because typically neural networks have
the structure of the inter-layer propagation which is naturally interpreted as a first order
differential equation when the continuum limit of the layers is taken\footnote{Highway networks \cite{highway}
may be another way.}.

The decoupling among the non-normalizable and the normalizable modes in the AdS/CFT correspondence
with the bulk scalar field $\phi$
works only for the free case, when the interaction vanishes, $V[\phi]=0$. 
%So, in the bulk, the scalar field should not have any interaction.
This is possible at the linear response level of the AdS/CFT correspondence.
In this appendix we use the coordinate system where the 
the AdS radial coordinate $\eta$ is a proper coordinate,
\begin{align}
ds^2 = -f(\eta) dt^2 + d\eta^2 + g(\eta)(dx_1^2 + \cdots + dx_{d-1}^2),
\end{align}
and define $h(\eta)\equiv \partial_\eta \log \sqrt{f(\eta) g(\eta)^{d-1}}$.
Then the free stationary scalar field equation is
\begin{align}
\left[\partial_\eta^2 + h(\eta)\partial_\eta +(\Box-m^2)\right]\phi = 0,
\label{sceq2}
\end{align}
where $\Box\equiv g(\eta)^{-2}\sum_{i=1}^{d-1}\partial_i^2$ is a covariant Laplacian.
As opposed to the case in \cite{Hashimoto:2018ftp,Hashimoto:2018bnb}, here we
included the spatial $(x)$ dependence of the external field and the response.
% in this formulation
%(although this is possible also for our first deep network of Fig.~\ref{fig:ournn} as well).
According to the holographic renormalization group \cite{deHaro:2000vlm}
%deHaro:2000vlm,Skenderis:2002wp,Papadimitriou:2004ap,Papadimitriou:2004rz,Papadimitriou:2010as}, 
the equation \eqref{sceq2}
can be rewritten as
\begin{align}
\left[\partial_{\eta} - b\right] 
\left[\partial_{\eta} - a\right] \phi = 0\, ,
\end{align}
with 
\begin{align}
& b(\eta,-\Box)+a(\eta,-\Box)=-h(\eta), \\
& b(\eta,-\Box)a(\eta,-\Box)-\partial_\eta a(\eta,-\Box)=\Box-m^2. 
\end{align}
These constraint equations are simply
\begin{align}
a^2 +h(\eta)a+\partial_\eta a=-\Box+m^2\, .
\label{aeq}
\end{align}
This generically allows two solutions $a(-\Box) \equiv f_{\pm}(-\Box)$, 
and with each of them, the scalar field equation reduces to a first order differential equation in $\eta$,
\begin{align}
\left[\partial_{\eta} - f_\pm(\eta,-\Box)\right] \phi = 0\, .
\end{align}
These two equations with $f_\pm$
govern
the non-normalizable and the normalizable modes, respectively. 
Therefore, once a spacetime bulk metric is given, we find two functions $f_\pm(-\Box)$, and use 
them to define the neural network by discretizing the $\eta$ direction as
$\eta = n \Delta \eta$.
Noting that the discretization of $\eta$ gives
\begin{align}
\partial_\eta \phi(\eta,x) = \frac{\phi(\eta + \Delta \eta)-\phi(\eta)}{\Delta \eta},
\end{align}
and 
the discretized spatial dependence
is interpreted as a convolution in the neural network,
\begin{align}
\partial_i^2 \phi(x_i)= \frac{\phi(x_i+\Delta x)-2\phi(x_i)+\phi(x_i-\Delta x)}{(\Delta x)^2},
\label{conv}
\end{align}
then we find that the neural network is defined
\begin{align}
\phi^{(n+1)}(x_i) = W_{\pm}^{(n)}\phi^{(n)}(x_i) 
\end{align}
where the network weights are 
\begin{align}
W_{\pm}^{(n)} = 1 + \Delta \eta \,f_\pm (\eta^{(n)},-\Box) .
\label{Wcon}
\end{align}

Note that we also discretize the $(d-1)$-dimensional space of the boundary QFT, as in \eqref{conv},
where the covariant Laplacian $\Box$ can be identified as a convolution in neural network.
Generically, spatial derivatives in field equations are identified as a combination of weights 
connecting nearby units. The locality of the bulk field theory is a constraint of the weights of the
neural networks.
In this way, we can always include spatial dependence 
of the external field and the response, as a convolutional neural network. 
So, \eqref{Wcon} defines a convolutional neural network equivalent to \eqref{sceq2},
with a trivial activation function. 

The left hand side of the neural network is governed by the propagation weight \eqref{Wcon}
for the non-normalizable mode. The input data $J(\vec{x})$ is placed at the initial layer $\eta =\eta_{\rm ini}\sim \infty$.
It propagates with $W_{+}^{(n)}$ toward the black hole horizon $\eta=0$, and is transformed to a
data $\phi_+(\vec{x},\eta=0)$. 
At the black hole horizon $\eta=0$, we need to impose the boundary condition $\partial_\eta \phi=0$.
Generically $\phi_+(\vec{x},\eta=0)$ does not satisfy it, so we need to complement it as
\begin{align}
\partial_\eta \left[
\phi_+(\vec{x},\eta) + \phi_-(\vec{x},\eta)\right]_{\eta=0}=0 \, .
\end{align}
This defines the initial condition for the right hand side of the neural network, 
$\phi_-(\vec{x},\eta)$, which propagates toward $\eta=\infty$ with $W_{-}^{(n)}$. Then we identify
the output $\phi_-(\vec{x},\eta=\infty)$ as $\langle {\cal O}(\vec{x})\rangle$.
We call this whole network, shown in Fig.\ref{fig:ae}, a holographic autoencoder. 

In reality for the training, 
we may focus on slowly varying external field and use the low momentum expansion 
$f_\pm = c_\pm(\eta) + d_\pm(\eta) \Box +\cdots $,  
then the weights are given as
\begin{align}
W_{\pm}^{(n)} = 1 + \Delta \eta \,\left(c_\pm(\eta^{(n)}) + d_\pm(\eta^{(n)}) \Box\right) \,.
\end{align}
We train the coefficient function $c_\pm$ and $d_\pm$ with a constraint that both 
consistently solve \eqref{aeq}. 

Using the horizon behavior $h(\eta)\sim 1/\eta$ and $g(\eta)\sim $const.,
we find that \eqref{aeq} has a universal solution
\begin{align}
c_{\pm} \sim \eta \, m^2/2, \quad d_{\pm} \sim -\eta \, \Box/2. 
\end{align}
This means that effectively the weight $W$ near the black hole horizon vanishes (except for the trivial 
``1+" part), due to the red shift factor via $h(\eta)$. Therefore, the effective dimensions of
the data space around the central part of the holographic autoencoder decrease, which is
suitable for the name ``autoencoder" usually used in machine learning.

\end{document}